\documentclass[a4paper,fleqn]{cas-dc}

\usepackage[numbers]{natbib}
\usepackage{xcolor} 
\usepackage{mhchem}
\usepackage{booktabs} 
\usepackage{siunitx}
\def\tsc#1{\csdef{#1}{\textsc{\lowercase{#1}}\xspace}}
\tsc{WGM}
\tsc{QE}
\tsc{EP}
\tsc{PMS}
\tsc{BEC}
\tsc{DE}
\begin{document}
%\linenumbers

\let\WriteBookmarks\relax
\def\floatpagepagefraction{1}
\def\textpagefraction{.001}

\shorttitle{The Fe$X$F ($X$ = O, S) monolayer}    

\shortauthors{X. Zhang, B. Wang, Y. Ge, Y. Liu, and W. Wan}  

\title[mode = title]{Near-room-temperature antiferromagnetism in Janus Fe$X$F ($X$ = O, S) monolayers}  

%\tnotemark[<tnote number>] 

%\tnotetext[<tnote number>]{<tnote text>} 



\author[1]{Xixiang Zhang}[orcid=0009-0001-9097-6036]
\credit{Methodology, Software, Formal analysis, Writing - Original Draft}

\author[1]{Busheng Wang}
\credit{Software}

\author[1]{Yanfeng Ge}[orcid=0000-0002-2441-7295]
\credit{Writing - review $\&$ editing, Funding acquisition}

\author[1]{Yong Liu}[orcid=0000-0002-5435-9217]
\credit{Writing - review $\&$ editing, Funding acquisition}

\author[1]{Wenhui Wan}[orcid=0000-0002-6824-0495]
\credit{Conceptualization, Funding acquisition, Review \& Editing, Supervision}
\cortext[1]{Corresponding author}
\cormark[1]
\ead{wwh@ysu.edu.cn}

\affiliation[1]{organization={State Key Laboratory of Metastable Materials Science and Technology $\&$ Hebei Key Laboratory of Microstructural Material Physics, School of Science, Yanshan University.},
	%          citysep={}, % Uncomment if no comma needed between city and postcode
	postcode={066004}, 
	country={People’s Republic of China}}
    
%\cortext[1]{Corresponding author}



\begin{abstract}
Inspired by the recently synthesized hexagonal layered phase of FeF$_2$, we studied the magnetic properties of the 1T-FeF$_2$ monolayer and its Janus Fe$X$F ($X$ = O, S) derivatives by first-principles calculations. Our results confirm that these materials are antiferromagnetic semiconductors, and that anion substitution effectively tunes their material properties: the band gap shifts from 3.37 eV (direct, FeF$_2$) to 2.35 eV (direct, FeOF) and 1.13 eV (indirect, FeSF); the magnetic moment of Fe ions increases; and the N\'{e}el temperature ($T_N$) rises dramatically to 248 K (FeSF) and 207 K (FeOF). Janus structures exhibit enhanced magnetic moment and direct AFM coupling. Under compression, $T_N$ is further optimized to 274 K ($-2$\% strain, FeSF) and 244 K ($-5$\% strain, FeOF). Both Janus materials retain their semiconducting nature and direction of easy magnetization axis under $\pm5$\% strain. This study validates the Janus structure as a viable approach to enhance 2D antiferromagnetism and highlights Fe-based oxyhalides as promising spintronic materials.
\end{abstract}


\begin{highlights}
\item The FeF$_2$ monolayer is an antiferromagnetic semiconductor with a low N\'{e}el temperature $T_N$ of 18 K.
\item Janus Fe$X$F ($X$ = O, S) monolayers exhibit enhanced magnetic moment and direct exchange coupling. 
\item The $T_N$ of FeSF monolayer reaches 274 K under a  biaxial compressive strain of -2\%. 
\item Janus Fe$X$F ($X$ = O, S) maintain robust semiconducting properties and stable easy axis under stress.
\end{highlights}

% Keywords
\begin{keywords}
 \sep Transition metal oxyhalides
 \sep Janus structure 
 \sep Antiferromagnetic semiconductor 
 \sep N\'{e}el temperature
 \sep Strain
\end{keywords}

\maketitle

% Main text
%\twocolumn
\section{Introduction}
Two-dimensional (2D) antiferromagnetic (AFM) materials exhibit an alternating alignment of magnetic moments, resulting in zero net magnetization. This configuration suppresses stray magnetic fields and enhances resilience against external magnetic disturbances, making 2D AFM systems promising candidates for next-generation spintronic and magnetic memory devices \cite{Olsen2024}. Meanwhile, 2D AFM materials exhibit intriguing phenomena such as giant magnetic exciton coupling \cite{shi2025giant}, magnon–phonon coupling \cite{sun2025strong}, and photothermal-electric effects \cite{zhou2025polarization}. However, experimentally, a limited number of 2D AFM materials have been synthesized, including FePS$_3$ \cite{wang2016raman,lee2016ising}, MnPS$_3$ \cite{Long2020}, MnPSe$_3$ \cite{Ni2021}, RuCl$_3$ \cite{Yang2023}, and NiI$_2$ \cite{Song2022}. The N\'{e}el temperature ($T_N$) of these 2D AFM materials is generally low (8--118K)~\cite{wang2016raman,lee2016ising,Long2020, Ni2021,Yang2023,Song2022}, 
while the technological relevance of 2D antiferromagnetism relies on magnetic order at room temperature. This highlights the need for further exploration and discovery of more 2D AFM materials to advance both fundamental research and device fabrication.

Previous theoretical studies have predicted a variety of 2D AFM materials with high $T_N$, such as the V$_{2}$C$X_{2}$ ($X$ = Cl, Br, I; $T_{N}$ = 420--590 K) \cite{luo2025prediction}, h-MnB ($T_{N}$ = 340 K) \cite{miao2024h}, CrMoC$_2$S$_6$ ($T_{N}$ = 556 K) \cite{Wang2022}, and Fe$_3$As monolayers ($T_{N}$ = 687 K) \cite{Yan2023}, which possess a magnetic anisotropic energy (MAE) between 24 and 1030 $\mu$eV per transition metal atom~\cite{luo2025prediction,miao2024h,Wang2022,Yan2023}. In contrast, for inherently weak AFM systems, significant research focus has been placed on enhancing their magnetic properties. Established methods include applying strain \cite{Wu2021}, carrier doping \cite{mi2022variation}, and introducing defects \cite{wang2018effect}. Given this context, Janus structure engineering has successfully enhanced ferromagnetism in 2D materials \cite{Wu2023,Ren2020}, but a similar strategy to improve 2D antiferromagnetism has not been reported.

The bulk FeF$_2$ crystallizes in the rutile structure (space group P4$_2$/mnm) and exhibits an AFM order with a $T_{N}$ of 78.4 K~\cite{PhysRevB.73.184434}.
This material has drawn considerable interest owing to its rich physical properties, including the formation of squeezed magnon states \cite{PhysRevB.73.184434}, magneto-optic coupling \cite{Lockwood2012}, spin–phonon interactions \cite{10.1063/1.342186}, perfect far-infrared reflection \cite{Ayuel2022}, and high capacity as an electrode in lithium-ion batteries \cite{Pereira2009,park2024entropy}.
Recently, Hao $\textit{et al.}$ reported pressure-induced phase transitions in bulk FeF$_2$ and synthesized a layered hexagonal phase (space group P-3m1) under pressure \cite{hao2024experimental}.
From this P-3m1 phase, a monolayer of 1T-FeF$_2$ can be exfoliated, benefiting from weak van der Waals (vdW) interlayer interactions.
Isostructural materials such as the 1T-Fe$X_2$ ($X$ = Cl, Br, I) monolayers are known to display a weak ferromagnetic (FM) order with low Curie temperatures of 42--109 K \cite{kulish2017single}. In contrast, the shorter Fe-Fe distance in the 1T-FeF$_2$ monolayer is a structural feature that generally favors AFM ordering. 
This tendency is consistent with the known effect of compressive strain in driving FM-to-AFM transitions in 2D materials \cite{Wu2025,10.1063/5.0055014,Li2021}. Despite these insights, the electronic and magnetic properties of the 1T-phase FeF$_2$ monolayer, as well as those of its Janus structures, remain unexplored.

In this work, using first-principles calculations, we systematically investigated the magnetic properties of FeF$_2$ and its Janus Fe$X$F (X = O, S) monolayers. The FeF$_2$ monolayer is an AFM semiconductor with a wide band gap of 3.49 eV but a low $T_N$ of 18 K. However, creating a Janus structure by substituting one layer of F with O or S dramatically enhances the AFM stability, 
yielding high $T_N$ values of 207 K for the FeOF monolayer and 248 K for the FeSF monolayer. Moreover, a compression of -2\% can increase the $T_N$ of the FeSF monolayer to 274 K. Meanwhile, the Janus structures can maintain its desirable semiconducting properties and the direction of easy magnetization axis under strain. Our results indicate the effective modulation of the AFM order by Janus engineering and the great potential of Janus transition-metal oxyhalides for room-temperature spintronics.

\begin{figure}[tb]
	\centering
	\includegraphics[width=0.45\textwidth]{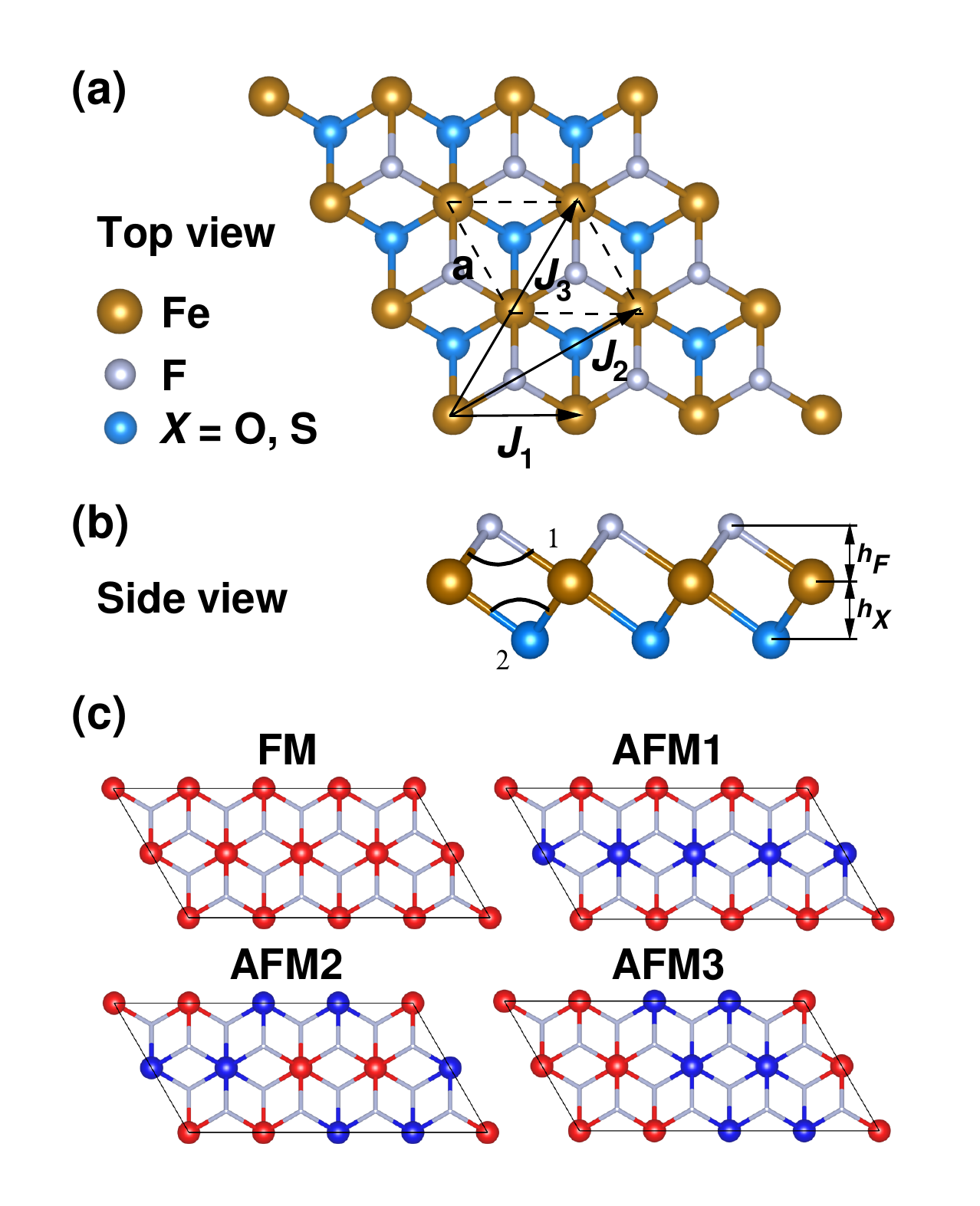}
	\caption{(a) The top and (b) side view of FeF$_2$ monolayer and Janus Fe$X$F ($X$ = O, S) monolayer. The brown, grey, and blue balls represent the Fe, F, and $X$ atoms, respectively. The primitive cell is labeled by bash lines. The exchange interaction $J_1$, $J_2$, and $J_3$ are displayed by arrows. (c) Four possible magnetic configurations, including FM, AFM1, AFM2, and AFM3 states. Here, the red and blue colors represent Fe atoms with up and down spins, respectively.}
	\label{fg1}
\end{figure}

\section{Computational method}
The first-principles calculations were performed using the Vienna ab
initio simulation package (VASP) \cite{Kresse1996} with projector-augmented wave (PAW) \cite{Bloechl1994} pseudopotentials and Perdew, Burke, and Ernzerhof (PBE) exchange-correlation functionals \cite{PhysRevLett.77.3865}.
The hybrid functional (HSE06) is adopted for an accurate prediction of band gaps \cite{Ernzerhof1999}. We adopted a plane-wave cut-off energy of 550 eV and a Monkhorst-Pack k-mesh of $12\times12\times1$ for sampling the first Brillouin zone. A vacuum layer of 20 {\AA} was adopted to eliminate the artificial interaction between the periodic images. The convergence criterion is $10^{-6}$ eV for the total energy and $10^{-2}$ eV/{\AA} for the atomic force. 
To account for the correlation effect of Fe-3$d$ electrons, we applied a simplified PBE+U approach which was proposed by Dudarev et al \cite{Dudarev1998}. An effective Hubbard term U = 4.0 eV was adopted for Fe-d orbitals according to previous studies \cite{yang2012structural,PhysRevB.109.014431}. 
We found that PBE+U and HSE06 yielded consistent magnetic ground state of FeF$_2$ monolayer [Figure S1]. The spin-orbit coupling (SOC) effect is included for magnetic anisotropy calculations. The phonon spectrums were calculated using a $4\times4\times1$ supercell, with the finite difference method as implemented in the Phonopy code \cite{10.1063/1.1564060,Togo2015}. Thermal stability was assessed via ab initio molecular dynamics (AIMD) simulations performed with a $4\times4\times1$ supercell, using a time step of 2 fs for a total duration of 6 ps. Monte Carlo (MC) simulations were used to estimate N\'{e}el temperature ($T_N$) with the $50 \times 50 \times \times 1$ supercell, using the MCsolver package \cite{Liu2019}.

\section{Results and discussion}
\subsection{FeF$_2$ monolayer}
Figures~\ref{fg1}(a) and~\ref{fg1}(b) illustrate the optimized crystal lattice of 1T-FeF$_2$ monolayer. The lattice constant and the Fe-F bind length are 3.26 \AA\ and 2.14 \AA, respectively. The local coordination reveals an octahedral environment around each Fe atom, bonded to six F atoms with a vertical Fe-F distance $h_{\rm F}$ of 1.01 \AA\ and a Fe-F-Fe bond angle of 99.54$^\circ$. Each F atom, in turn, bridges three Fe atoms. To determine the ground-state structure, we mapped the potential energy surface as a function of the fluorine position. The result [Fig. S2(a)] exhibits a single minimum located above the hexagonal hollow site, which corresponds to the 1T phase.

\begin{table*}[htbp]
	\centering
	\caption{Lattice constant ($a$), bond lengths ($d_{\mathrm{Fe-F}}$, $d_{\mathrm{Fe-X}}$), In-F and In-O vertical distances ($h_F$, $h_X$), bond angles ($\alpha_1$, $\alpha_2$), elastic constants ($C_{11}$, $C_{12}$, $C_{66}$), Young's modulus ($Y$), and Poisson's ratio ($\nu$) of $\mathrm{FeF_2}$, $\mathrm{FeOF}$, and $\mathrm{FeSF}$ monolayers.} 
	\label{tab:1} 
	\begin{tabular}{lccccccccccccc}
		\toprule 
		Materials & $a$ & $d_\text{Fe-F}$ & $d_\text{Fe-$X$}$ & $h_{\rm F}$ & $h_X$ & $\alpha_1$ & $\alpha_2$ & $C_{11}$ & $C_{12}$ & $C_{66}$ & $Y$ & $\nu$  \\
		\cmidrule(lr){2-6} \cmidrule(lr){7-8} \cmidrule(lr){9-12} \cmidrule(lr){13-13}
		& \multicolumn{5}{c}{(\AA)} & \multicolumn{2}{c}{($^\circ$)} & \multicolumn{4}{c}{(N/m)} & \\
		\midrule 
		FeF$_2$  & 3.26 & 2.14 & 2.14 & 1.01 & 1.01 & 99.54 & 99.54 & 33.42 & 18.97 & 7.22 & 22.65 & 0.57 \\
		FeOF     & 3.08 & 2.15 & 1.94 & 1.20 & 0.77 & 91.81 & 105.39 & 77.34 & 39.00 & 19.17 & 57.66 & 0.50 \\
		FeSF     & 3.35 & 2.20 & 2.38 & 1.06 & 1.39 & 98.93 & 89.40 & 31.24 & 19.69 & 5.77 & 18.82 & 0.63 \\
		\bottomrule
	\end{tabular}
\end{table*}

\begin{figure*}[bthp!]
	\centering
	\includegraphics[width=0.9\textwidth]{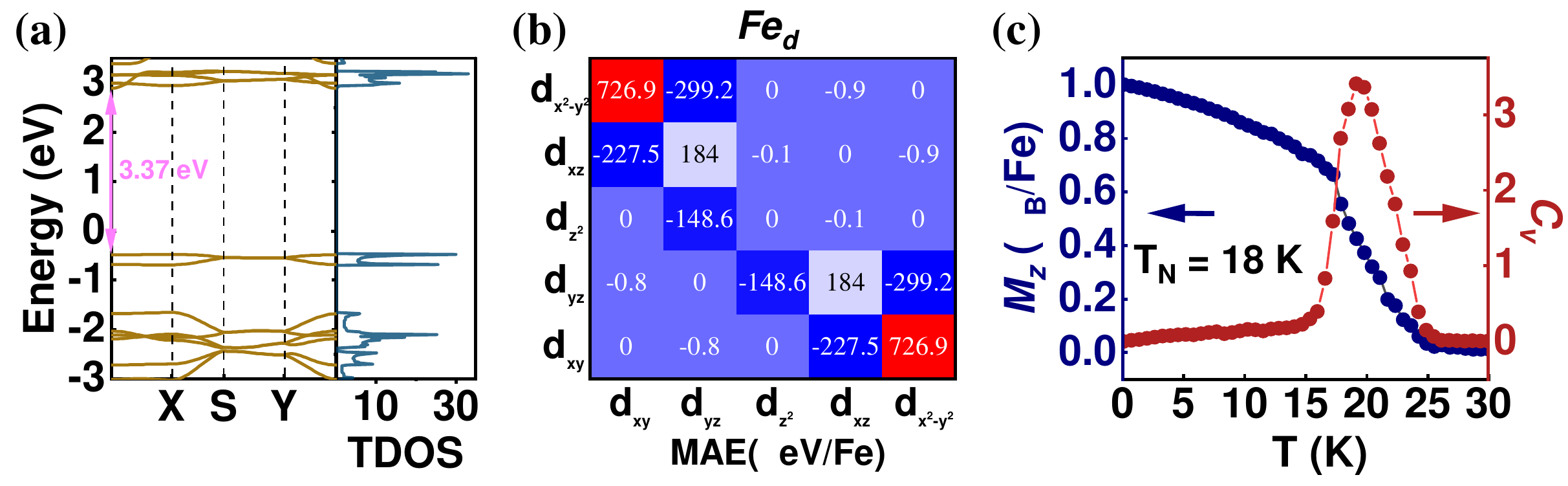}
	\caption{(a) Band structure and total density of states (TDOS) of the FeF$_2$ monolayer. (b) Orbital-projected contribution of Fe-$d$ orbitals to the MAE. (c) The normalized magnetic moment ($M_z$) and heat capacity ($C_v$) of Fe atoms as function of temperature, which is obtained through the MC simulation.}
	\label{fg2}
\end{figure*}

The FeF$_2$ monolayer exhibits excellent structural stability across multiple criteria. 
First, its dynamic stability is verified by the absence of imaginary frequencies in the phonon dispersion [Fig. S3(a)]. Second, its thermal stability at room temperature is confirmed by AIMD simulations, which show only minor structural fluctuations [Fig. S3(b)]. Finally, its mechanical stability is established as the elastic constants comply with the Born criteria for a hexagonal lattice ($\textit{C}_{11}$>0, $\textit{C}_{66}$ > 0 and $\textit{C}_{11}$$\textit{C}_{22}$ >$\textit{C}_{12}^2$) [Table 1]. The cohesive energy of the FeF$_2$ monolayer is defined as 
\begin{equation}
	\begin{aligned}
		E_{coh} &= E({\rm FeF_2})-E({\rm Fe})-2E({\rm F}),
		\label{coh}
	\end{aligned}
\end{equation}
where $E(\rm FeF_2)$, $E(\rm Fe)$, and $E(\rm F)$ are the energies of the FeF$_2$ monolayer, isolated Fe atom, and isolated F atom, respectively. 
The calculated $E_{coh} = - 4.10$ eV for the FeF$_2$ monolayer is comparable to the values reported for other stable 2D materials such as MoI$_3$ ($-4.35$ eV/atom) \cite{PhysRevB.103.075433}, Fe$_2$Si ($-4.10$ eV/atom) \cite{sun2017room}, and FeB$_2$ ($-4.87$ eV/atom) \cite{zhang2016dirac} monolayers, further confirming its structural robustness.  

We determined the magnetic ground state of the FeF$_2$ monolayer by comparing the total energies of four candidate configurations: FM, AFM1, AFM2, and AFM3 [Fig.~\ref{fg1}(c)]. 
Our total energy calculations identify the AFM1 state as the magnetic ground state [Table S1]. Electron Localized Function (ELF) analysis reveals electron transfer from Fe to F ions [Fig. S4(a)], indicative of an ionic interaction. In a purely ionic picture, the Fe$^{2+}$ ion ($3d^6$ configuration) would possess a localized magnetic moment of 4 $\mu_B$. The calculated magnetic moment of 3.81 $\mu_B$ per Fe ion suggests a deviation from this ideal, implying the presence of a minor covalent component in the Fe-F bonds.

We calculated the band structure of the FeF$_{2}$ monolayer in its ground state AFM1, using the primitive cell [Fig. S5(a)]. The FeF$_{2}$ monolayer exhibits a wide direct band gap of 3.37 eV at the $\Gamma$ point [Fig.~\ref{fg2}(a)], which is notably larger than the bulk value (2.57 eV) \cite{yang2012structural}. Flat bands near both valence band maximum (VBM) and the conduction band minimum (CBM) suggest a large carrier effective mass. The projected density of states (PDOS) analysis [Fig. S6(a)] shows that the VBM is primarily composed of Fe $d_{xz}$ and $d_{x^2-y^2}$ orbitals, while the CBM is dominated by Fe $d_{xy}$ and $d_{yz}$ orbitals.

A sizable MAE is crucial for stabilizing magnetic order in 2D materials against thermal fluctuations. Our calculations reveal that the FeF$_2$ monolayer possesses an easy axis along the $x$ direction with a MAE of 235 $\mu$eV/Fe [Table. S1]. This value is higher than that of $h$-MnB (154 $\mu$eV/Mn) \cite{miao2024h} but smaller than the FeCN$_2$ monolayer (500 $\mu$eV/Fe) \cite{wang2023antiferromagnetic} and the Fe$_3$As monolayer (1.03 meV/Fe) \cite{Yan2023}.

To explain the in-plane magnetic anisotropy of the FeF$_2$ monolayer, we expressed the MAE 
within the framework of second-order perturbation theory: 
\begin{equation}
	\begin{aligned}
		\mathrm{MAE} &= E^{\mathrm{SOC}}_z - E^{\mathrm{SOC}}_x \\
		&= \xi^{2} \sum_{o, u} \frac{ 
			| \langle o | S_{z} L_{z} | u \rangle |^{2} - 
			| \langle o | S_{x} L_{x} | u \rangle |^{2} 
		}{ \varepsilon_{o} - \varepsilon_{u} }.
		\label{eq7}
	\end{aligned}
\end{equation}
Here, $E^{SOC}_z$ and $E^{SOC}_x$ represent the total energy with spins aligned along the $z$ and $x$ directions, respectively. 
Given that the denominator $\varepsilon_{u}-\varepsilon_{o}$ in Eq.~\ref{eq7} places greater weight on electronic states near the Fermi level, and considering that Fe-$d$ orbitals dominate the PDOS in this region [Fig. S6(a)], they are the primary contributors to the MAE of FeF$_2$ monolayer. We classified the $d$ orbitals by magnetic quantum number $m$. We used $d_{m=0}$, $d_{|m|=1}$, and $d_{|m|=2}$ to represent $d_{z^2}$, \{$d_{xz}$, $d_{yz}$\}, and \{$d_{xy},d_{x^2-y^2}$\}, respectively. Table S2 lists the values of $| \langle o | S_{z} L_{z} | u \rangle |^{2} - | \langle o | S_{x}  L_{x} | u \rangle |^{2}$. Within the same spin channel, $d$ orbitals contribute positively to the MAE by the matrix element $\left\langle {m} \right|{\hat L_z}\left| {m} \right\rangle $, but negatively via 
$\left\langle {{m}} \right|{\hat L_x}\left| {m\pm1} \right\rangle $ in Eq.~\ref{eq7}. The SOC effect between different spin states gives an opposite contribution to the MAE in the summation of Eq.~\ref{eq7}. The orbital-resolved MAE [Fig.~\ref{fg2}(b)] shows that the positive magnetic anisotropy in the FeF$_2$ monolayer primarily originates from the coupling of $\left\langle {d_{x^2-y^2}} \right|{\hat L_z}\left| {d_{xy}} \right\rangle$ within the same spin channel.

We adopted a Heisenberg model to estimate the N\'{e}el temperature ($T_N$) of FeF$_2$ monolayer:
\begin{equation}
	H=-\sum_{i, j} J_{ij} S_{i}\cdot S_{j}-D\sum_{i} (S_{i}^{e})^{2}. 
	\label{eq1}
\end{equation}
Here, $S_{i}$ is the spin vector at the $i$th site, $S_{i}^{e}$ is the spin component along the easy axis, and $D$ is the single-ion anisotropic energy. We considered exchange constant $J_{ij}$ up to the third-nearest neighbors ($J{_1}$, $J{_2}$, and $J{_3}$), as shown in Fig.~\ref{fg1}(a). With the magnetic moment normalized to 1, the energies of four magnetic states are given by:
\begin{equation}
	\begin{aligned}
	 E_{FM} &= E_0 - 3J_1 - 3J_2 - 3J_3 - D, & \\
	 E_{AFM1} &= E_0 + J_1 + J_2 - 3J_3 - D, & \\
	 E_{AFM2} &= E_0 + J_1 - J_2 + J_3 - D, & \\
	 E_{AFM3} &= E_0 - J_1+ J_2 + J_3 - D. &
	\end{aligned}
\end{equation}

Substituting the energies from the first-principles calculations into the above equations, the exchange interactions are estimated as $J_1$ = 0.19 meV, $J_2$ = $-4.00$ meV and $J_3$ = 2.01 meV. The large negative value $J_2$ indicates a strong AFM interaction between the nearest-neighboring Fe ions. MC simulations estimated that the $T_N$ of FeF$_2$ monolayer is only 18 K [Fig.~\ref{fg2}(c)]. 

The suppressed $T_N$ results from the competition between the FM interactions ($J_1$, $J_3$) and the AFM interaction $J_2$. The Fe-F-Fe bond angle between the nearest-neighboring Fe ions is 99.54$^\circ$.
According to the Goodenough-Kanamori-Anderson (GKA) rules \cite{Goodenough1958, PhysRev.115.2}, this angle, being close to 90°, favors a FM superexchange for $J_1$.
In contrast, the Fe-F-Fe bond angle between the second nearest-neighboring Fe ions is 136.8$^\circ$. Being closer to 180$^\circ$, this geometry is highly conducive to the AFM superexchange observed for $J_2$.

\subsection{Fe$X$F ($X =$ O and S) monolayer}
The low $T_N$ of the FeF$_2$ monolayer limits its potential for spintronics applications. 
To address this, we designed Janus structures by substituting one fluorine layer with oxygen group elements, aiming to enhance the antiferromagnetic ordering in the FeF$_2$ monolayer. Among the various configurations studied, phonon dispersion calculations [Fig. S7(a, b)] confirm two stable Janus monolayers: Fe$X$F ($X$ = O, S). The potential energy surface for both structures reveals a single minimum, corresponding to a stable hexagonal phase with no alternative metastable states [Fig. S2(b, c)]. The calculated cohesive energies of –4.55 and –3.38 eV/atom for FeOF and FeSF monolayers, respectively, suggest that these compounds are synthetically feasible. Furthermore, AIMD simulations at 300 K [Fig. S7(c, d)] and elastic constants [Table~\ref{tab:1}] demonstrate their robust thermodynamic stability and mechanical stability.

The in-plane mechanical properties of the Janus Fe$X$F ($X$ = O, S) monolayers, including the Young's modulus ($Y$) and Poisson's ratio ($\nu$), are isotropic, as observed in the parent FeF$_2$ monolayer [Figs. S7(e-h)]. As listed in Table~\ref{tab:1}, the FeOF monolayer exhibits larger elastic constants and a higher Young's modulus than FeF$_2$, whereas the values for the FeSF monolayer are comparable. The Young's moduli of the Janus Fe$X$F monolayers are notably lower than those of Phosphorene (140 N/m) \cite{wu2020stabilities} and MoS$_2$ monolayer (130 N/m) \cite{cooper2013nonlinear}, but higher than that of CrI$_3$ (28 N/m) \cite{zheng2018tunable}, indicating good flexibility and the ability to withstand external strain. In contrast, the Poisson's ratio ($\nu$) of Fe$X$F ($X$ = O, S) monolayers exceeds 0.33, suggesting ductile mechanical behavior according to the Frantsevich rule \cite{frantsevich1983elastic}.

\begin{figure*}
	\centering
	\includegraphics[width=1.0\textwidth]{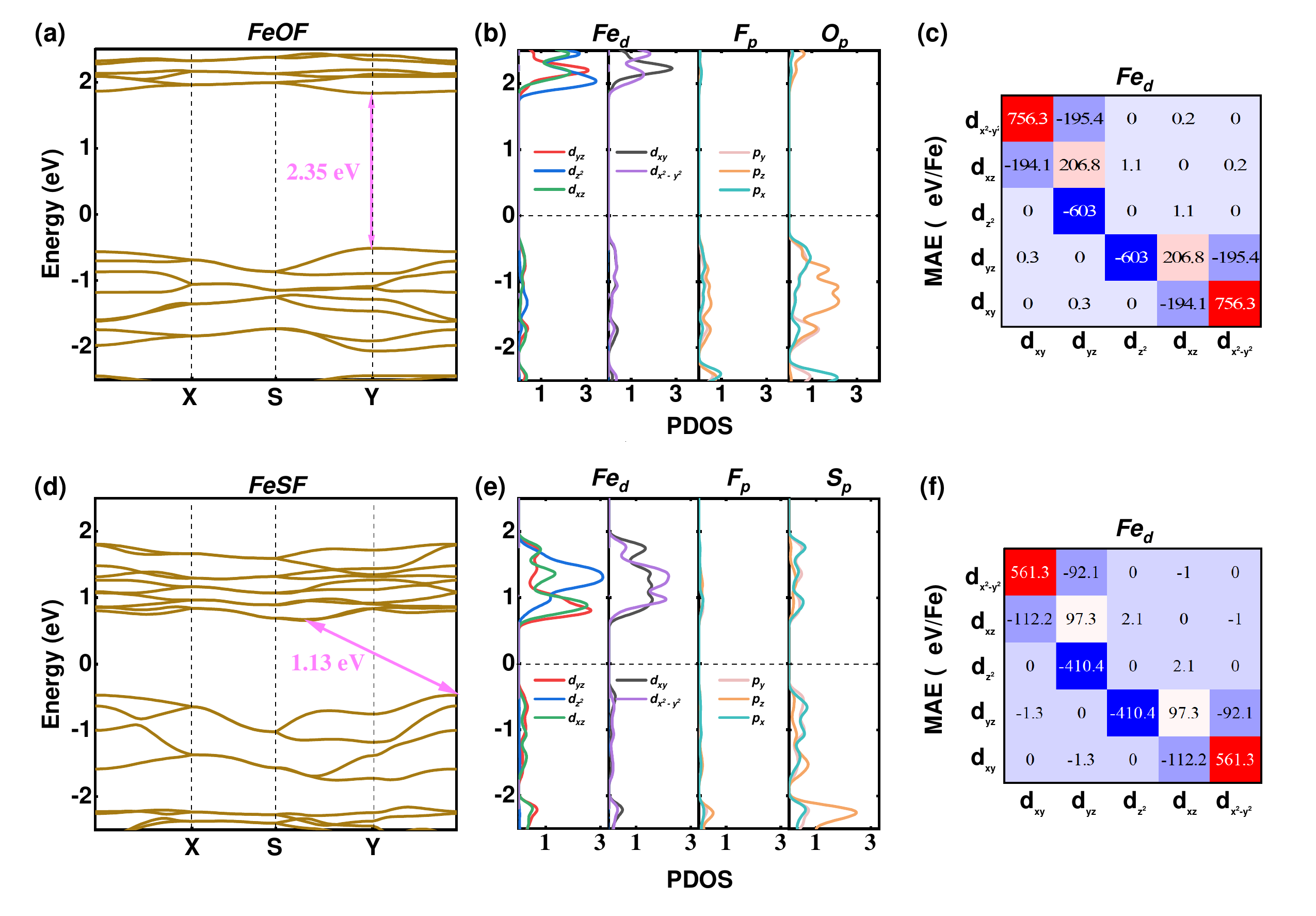}
	\caption{Band structures and density of states diagrams of (a) (b) FeOF and (d) (e) FeSF monolayers. Orbital-projected contribution to the MAE from the Fe-$d$ orbitals in (c) FeOF and (f) FeSF monolayers.}
	\label{fg3}
\end{figure*}

As shown in Table S1, the Janus Fe$X$F ($X$ = O, S) monolayers exhibit a different magnetic ground state from the FeF$_2$ monolayer. Although the latter has an AFM1 ground state, the former stabilizes in the AFM2 configuration. Furthermore, the energy difference between the FM and AFM states is significantly larger in the Janus structures, indicating that constructing Janus structure enhances the AFM order.

We calculated the band structure of the Fe$X$F ($X$ = O, S) monolayers in the AFM2 state using a primitive cell [Fig. S5(b)]. The FeOF monolayer is a direct-gap semiconductor with a band gap of 2.35 eV, while the FeSF monolayer exhibits an indirect gap of 1.13 eV [Fig.~\ref{fg3}(a) and~\ref{fg3}(d)]. Their CBMs is primarily composed of Fe-$d$ orbitals: specifically the $d_{z^2}$ orbital in FeOF monolayer and ($d_{yz}$, $d_{xz}$) orbitals in FeSF monolayer. In contrast, their VBM is mainly contributed by the $p$ orbitals of nonmetal atoms (O and S, respectively) [Fig.~\ref{fg3}(b) and~\ref{fg3}(e)]. At deeper energy levels, the $p$ orbitals of O, S, and F atom hybrids with Fe-$d$ orbitals across a wide energy range from -8 eV to the Fermi level [Figs. S6(b, c)]. 

A key difference between the FeF$_2$ and Fe$X$F ($X$ = O, S) monolayers is their lattice symmetry, which dictates their electronic structures. FeF$_2$ (AFM1) retains inversion symmetry, while Fe$X$F (AFM2) does not [Fig. S5]. 
This broken symmetry results from substituting half of the F$^{-}$ ions with the O$^{2-}$ or S$^{2-}$ ions; their larger atomic radius and different electronegativity introduce an asymmetric elongation distortion along $z$ axis. 
This distortion thereby lifts the degeneracy of the Fe-$d$ orbitals, consequently forming the highly dispersive band structures presented in Fig.~\ref{fg3}(a, d).  

Substituting F atoms with $X$ atoms enhances the magnetic moment of Fe ions in Janus Fe$X$F ($X$ = O, S) monolayers, which we attribute to the modified Fe-$X$ bonding. The ELF in Fig. S4(b) provides direct evidence: the more extensive localized electron regions near the O or S atoms indicate stronger charge transfer compared to FeF$_2$. The consequent electron accumulation around these anions increases the local moment on Fe. In an ideal ionic bonding, Fe$^{3+}$ ($3d^5$) would yield a moment of 5 $\mu_B$. Our computed values of 4.28 $\mu_B$ (FeOF) and 4.01 $\mu_B$ (FeSF) exceed the 3.81 $\mu_B$ moment of Fe$^{2+}$ in FeF$_2$ but fall short of the full ionic value. This deviation confirms the presence of partial electron sharing in the Fe–$X$ bond, underscoring its more covalent character relative to the Fe–F bond.

The MAE values of the Fe$X$F monolayers are -37 (FeOF) and 52 (FeSF) $\mu$eV/Fe [Table. S1]. The positive and negative values of MAE respectively indicate the easy magnetization axes is in-plane and out-of-plane.
The suppressed MAE magnitude is explained by the downshift of Fe-$d$ orbitals in PDOS [Fig. S6(b, c)], which increases the ($\varepsilon_{u}-\varepsilon_{o}$) energy separation in the denominator of Eq.~\ref{eq7}. Comparatively, the MAE of Fe$X$F monolayer is comparable to that of the Fe$_2$C monolayer (20 $\mu$eV/Fe) \cite{Lou2022} and the FeOCl monolayer (50 $\mu$eV/Fe) \cite{wang2020}, but is lower than that of the FeF$_2$ monolayer (235 $\mu$eV/Fe) and CrI$_3$ monolayer (500 $\mu$eV/Cr) \cite{kumar2019magnetism}. The orbital-resolved analysis in Fig.~\ref{fg3}(c) identifies the origin of the negative MAE of the FeOF monolayer: it arises mainly from the substantial negative contribution of the orbital coupling $\left\langle {d_{z^2}} \right|{\hat L_x}\left| {d_{yz}} \right\rangle$ between the d$_{z^2}$ and $d_{yz}$, driven by the high spectral weight of the d$_{z^2}$ orbital near the CBM and the wide distribution of $d_{yz}$ in the valence bands [Fig. S6(b)]. The positive MAE in the FeSF monolayer shows a similar origin to that of the FeF$_2$ monolayer, both primarily originates a coupling of $\left\langle {d_{x^2-y^2}} \right|{\hat L_z}\left| {d_{xy}} \right\rangle$ within the same spin channel [Fig.~\ref{fg3}(f)].

The distinct magnetic anisotropy in FeOF and FeSF monolayers originates from the ligand-modulated competition between spin-orbit coupling (SOC) channels. The smaller lattice constant of the FeOF monolayer imposes a compressed crystal field, under which the strong negative contribution from the coupling between the $d_{yz}$ and $d_{z^2}$ orbitals (-603 $\mu$eV/Fe) prevails over the net positive contributions, stabilizing a perpendicular easy axis. Replacing O with S significantly weakens this key $d_{yz}$-$d_{z^2}$ coupling (-410 $\mu$eV/Fe). This effect is compounded by a concurrent reduction in other negative channels, which collectively diminishes the total negative weight and thereby allows the positive contributions to dominate the MAE balance. Consequently, the easy axis switches to in-plane.

In Janus Fe$X$F ($X$ = O, S) monolayers, the diminished MAE is compensated for by their enhanced antiferromagnetism. 
Computed exchange constants confirm this: ($J_1$, $J_2$, $J_3$) are ($-32.65$, $-0.46$, $-1.09$) meV for the FeOF monolayer and ($-35.81$, $-4.38$, $-12.89$) meV for the FeSF monolayer. In the FeOF monolayer, strong Fe-O electrostatic attraction shrinks the lattice constants (Fe-Fe distance) and the bond length compared to the FeF$_2$ monolayer.  
As shown in Table~\ref{tab:1}, the Fe-Fe distance in the FeOF monolayer is 3.08 \AA\, which is smaller than that of the FeF$_2$ monolayer. The overlap of adjacent $d$-shells of Fe ions strengthens the direct AFM coupling ($J_1$). 
Meanwhile, the direct exchange coupling between the half-filled and filled $d$-shell tends to be AFM and FM, respectively \cite{Goodenough1963}. 
The Fe$^{3+}$ state also favors AFM alignment due to fewer valence electrons than Fe$^{2+}$ state.
In the FeSF monolayer, despite the larger lattice lattice caused by the larger atomic radius of S, the Fe-$d$/S-$p$ hybridization redistributes orbital occupancy of the Fe-$d$ orbitals. As shown in Fig. S8, the occupation in $d_{z^2}$ decreases and other $d$ orbitals increase in the FeSF monolayer, increasing in-plane Fe-$d$ wavefunction overlap and the direct $d \leftrightarrow d$ exchange $V_{dd\sigma}$ in FeSF monolayer \cite{Huang2019}. That also enhances $J_1$ in FeSF monolayer. 
On the other hand, the Fe-$X$-Fe bond angles ($\sim$90$^\circ$) are similar to the FeF$_2$ monolayer [Table~\ref{tab:1}]. According to the GKA rules \cite{Goodenough1958, PhysRev.115.2}, the Janus structure negligibly affects the FM-type super-exchange coupling between Fe ions. Thus, the enhanced antiferromagnetism in both Fe$X$F monolayers primarily stems from the strengthened direct AFM coupling ($J_1$).

\begin{figure*}
	\centering
	\includegraphics[width=1.0\textwidth]{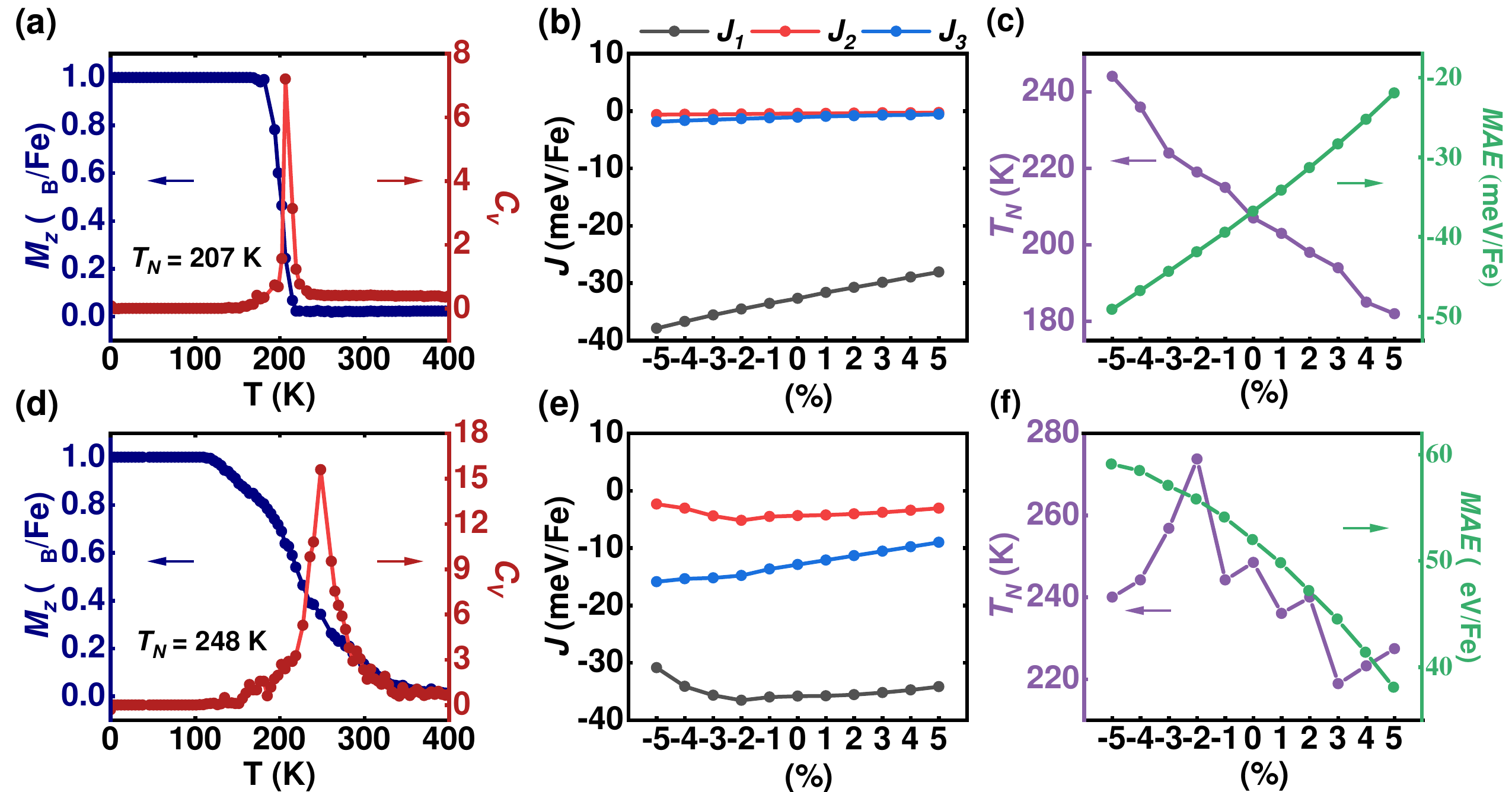}
\caption{
(a) (d) Normalized magnetic moment ($M_Z$) and heat capacity ($C_v$) versus temperature; 
(b) (e) Evolution of exchange constants ($J_1$, $J_2$, $J_3$) under equibiaxial strain from $-5\text{\%}$ to $5\text{\%}$; 
(c) (f) Strain dependence of N\'{e}el temperature ($T_{\mathrm{N}}$) and magnetocrystalline anisotropy energy (MAE). (a-c) correspond to FeOF monolayer, (d-f) to FeSF monolayer.}
	\label{fg4}
\end{figure*}

With a large AFM-type $J_1$, the $T_N$ for Fe$X$F ($X$ = O and S) monolayers reach 207 K and 248 K, respectively [Figs.~\ref{fg4}(a) and~\ref{fg4}(d)], which is much larger than that of the FeF$_2$ monolayer (18 K), FePS$_3$ monolayer (118 K) \cite{lee2016ising}, and NiPS$_3$ bilayer (130 K) \cite{kim2019suppression}. 

We investigate the influence of biaxial strain (from -5\% to +5\%) on the magnetic properties of Janus Fe$X$F ($X$ = O and S) monolayers, a known tuning method for 2D materials \cite{lv2019strain, cenker2022reversible}. Notably, the AFM2 magnetic ground state and semiconducting behavior are preserved across this range, as detailed in Tables S3 and S4. This stability against strain is a key requirement for ensuring the operational reliability of related spintronic devices.

The magnetic proprties of FeOF and FeSF monolayers exhibit distinct responses to external strain. 
Under compression, the key exchange parameter $|J_1|$ exhibits a divergent strain dependence: 
it strengthens monotonically in the FeOF monolayer but undergoes an anomalous suppression in the FeSF monolayer beyond -2\% strain, as shown in Fig.~\ref{fg4}(b) and \ref{fg4}(e).

It is worth noting that the MAE varies relatively little within the ±5\% strain range while maintaining the original direction of easy magnetization axis, which ensures stability for spintronic device. The evolution trends of the MAE of the FeOF and FeSF monolayers under strain are different. The band gap of the FeOF monolayer increases under compressive strains [Fig. S9(e)], and maintain the direct band gap [Fig. S9(a)].
The enlarged band gap increases the $\varepsilon_{u}-\varepsilon_{o}$ in Eq.~\ref{eq7} and thus decreases the MAE [Fig.~\ref{fg4}(c)]. On the other hand, tensile strains decrease the band gap and raise the MAE of the FeOF monolayer. Moreover, when the material is stretched by 5\%, the valence band top begins to shift towards the $\Gamma$ point, forming an indirect band gap [Fig. S9(c)].

The band gap of the FeSF monolayer exhibits an opposite trend under strains [Fig. S9(f)]. The band gap continuously decreases with the increase in compression. 
Thus, the MAE of FeSF monolayer increases under compression [Fig.~\ref{fg4}(f)].
In contrast, under tension, its behavior is different from that of the FeOF monolayer. At a 1\% tensile strain, the band gap reaches its maximum value (1.131 eV), and then it continues to decrease. The FeSF monolayer maintains the indirect bandgap property within the strain range of -5\% to 5\% [Fig. S9(b) and S9(d)]. 

The regulation of $T_N$ of the Fe$X$F monolayer by strain shows a material dependence. For the FeOF monolayer, $T_N$ monotonically increases with compressive strain increasing and reaches 244 K at a strain of -5\% [Fig.~\ref{fg4}(c)]. In contrast. The $T_N$ of the FeSF monolayer reaches a peak of 274 K at $-2\%$ compression [Fig.~\ref{fg4}(f)], then decreases with further strain deviation (dropping to 227 K at $+5\%$ and 240 K at $-5\%$ strain). 
This behavior can be attributed to the strain response of $J_1$ --- The dominant mechanism $T_N \propto |J_1|$ holds for FeOF and FeSF monolayers.

\section{Conclusions}
In summary, based on first-principles calculations, we demonstrate that the hexagonal 1T-FeF$_2$ monolayer is an AFM semiconductor with a direct band gap of 3.37 eV and a low $T_N$ of 18 K. Substituting F anions with oxygen or sulfur leads to the formation of stable Janus Fe$X$F ($X$ = O, S) monolayers. The resulting FeOF and FeSF monolayers exhibit a direct band gap of 2.35 eV and an indirect band gap of 1.13 eV, respectively. In both Fe$X$F systems, the Fe$^{3+}$ ions carry a large magnetic moment, due to enhanced charge transfer between Fe and O or S ions. Notably, the direct AFM coupling between Fe ions is strengthened, owing to the reduced Fe-Fe distance and increased occupation of in-plane Fe-$d$ orbitals. Consequently, $T_N$ is enhanced by an order of magnitude, reaching 207 K for FeOF and 248 K for FeSF. Under  compression, $T_N$ further increases to 244 K for FeOF at $-5$\% strain and to 274 K for FeSF at $-2$\% strain. Moreover, both Fe$X$F systems retain robust semiconducting behavior, magnetic ground state, and easy magnetization axis under strains ranging from $-5$\% to 5\%. These findings establish Janus engineering as an effective strategy for enhancing 2D antiferromagnetism. Our study also suggests that 2D Fe-based oxyhalides are promising candidate materials for AFM spintronic applications.

%\indent\textcolor{blue}{\em Conclusion.}---

% Uncomment and use as the case may be
%\begin{theorem} 
%\end{theorem}

% Uncomment and use as the case may be
%\begin{lemma} 
%\end{lemma}


%\section*{}\label{}

\printcredits

\section*{Declaration of competing interest }
The authors declare no conflict of interest.

\section*{Data Availability}
Data will be made available on request

\section*{Acknowledgements}
This work was supported by National Natural Science Foundation of China (No. 11904313), the Project of Hebei Educational Department, China (No. BJ2020015),
Cultivation Project for Basic Research and Innovation of Yanshan University (No.2022LGZD001), and the Innovation Capability Improvement Project of Hebei province (Grant No. 22567605H). The numerical calculations have been done in the High Performance Computing Center of Yanshan University.

\section*{Supplemental Material}
See Supplemental Material for further details, including comparison of the ground state of Fe$X$F ($X$ = O, F) monolayer calculated by HSE06 and PBE methods, stability of the structure, the electronic structures of the FeF$_2$ and Fe$X$F monolayers, and the data related to strain. 

%\bibliographystyle{cas-model2-names}
\bibliographystyle{rsc}

\bibliography{rsc}



\end{document}